\documentclass[12pt]{article}
\usepackage[dvips]{graphicx}
\usepackage{latexsym}
\usepackage[affil-it]{authblk}
\usepackage{amsmath}
\usepackage{amssymb}
\usepackage{color}

\newcommand{\wt}[1]{\tilde{#1} }

\begin{document}
\title{Planar Black holes and Entanglement Entropy in Analog Gravity Models}

\author{Neven Bili\'c$^1$\thanks{bilic@irb.hr} and Tobias Zingg$^{2}$}

\affil{$^1$Division of Theoretical Physics, Rudjer Bo\v skovi\'c Institute \\ 10002 Zagreb, Croatia\\
$^{2}$Department of Physics and Helsinki Institute of Physics\\
P.O.Box 64, FIN-00014 University of Helsinki, Finland	
}

\maketitle


\begin{abstract}
Via constructing an explicit Lagrangian for which the perturbation equations 
are analogues of a scalar field propagating in a planar black hole space-time, 
it is found that all planar black holes conformal to a Painlev\'e--Gullstrand type line element can be realized as analogue metrics.
 We also introduce the concept of  holographic entanglement entropy for planar black-hole space-times. 
This is valid for an arbitrary choice 
of conformal and blackening factor, thereby vastly extending the number of known examples of explicitly 
known analogue metrics. 
\end{abstract}


\thispagestyle{empty}


%

\section{Introduction}
\label{sec:intro}
Certain condensed matter systems, 
respectively, the Lagrangians describing these have a property that small perturbations around a given background are described by the equations of motion of a field propagating in curved space-time. 
Thus, these systems may serve as `analogues' of phenomena in gravitational physics and could,  in principle, be employed to simulate gravity in tabletop experiments.
Though already known in theory since the 1980s \cite{Unruh:1980cg, Barcelo2011},  only in recent years has this approach to simulating  gravity  attracted more attention,
mostly due to new technologies -- in particular, in dealing with Bose-Einstein condensates 
or cold atom systems -- having been 
developed and making these kinds of experiments more 
accessible 
\cite{Fedichev:2003id,Weinfurtner:2010nu,Steinhauer:2014dra,Steinhauer:2015saa,Euve:2015vml,Peloquin:2015rnl,deNova:2018rld}.

However, it should be noted that, a priori, not all interesting geometries can be mimicked by analogue geometries. 
For example, counting the available degrees of freedom, in general relativity (GR) in $3+1$ dimensions, we 
have four degrees of freedom per point in space-time: $10$ independent components of a general metric minus 
$6$ owing to the $6$ independent Einstein equations. 
Whereas an analogue metric basically depends on two independent functions, 
which are the scalar potential $\theta$ that generates the flow velocities and the speed of sound $c$. 
Thus, the basic analogue gravity setup involving a single scalar field, which is also what we will consider in the following, can not reproduce all possible metrics that could be derived from GR.
However, as additional degrees of freedom enter by coupling to an external potential, which is assumed to be freely tunable, the setup considered is actually more than sufficient to mimic the most important phenomena, such as black holes, FRW cosmology, and even some aspects of semiclassical quantum gravity, such as Hawking radiation. 

As these phenomena are all of central importance in gravity physics, it is desirable to extend the class of analog gravity systems to as many metrics as possible.
Besides astronomic observations, analog gravity provides the only way to experimentally test such predictions in a lab environment.
In this paper, we follow the formalism developed in \cite{Bilic:2018fsk} where it was restricted to a  planar AdS$_5$ black hole (BH), 
and extend the applicability of analog gravity by demonstrating that it can potentially capture all phenomena described by a field propagating
in any space-time that is conformal to a rather generic stationary planar BH, for an arbitrary choice of blackening factor.
Our paper provides a generalization and adds to the examples of planar space-times that have already been found to have an analog dual \cite{Hossenfelder:2014gwa,Hossenfelder:2015pza,Dey:2016khw}.

We aim to generalize possible analog planar geometric structures
that could, in principle, be accomplished by a suitable design of the fluid flow.
In analog gravity, the planar BHs may appear if a fluid flows along one coordinate dimension so that two other space dimensions become irrelevant. The reasons for considering primarily planar black holes are threefold.

First, black holes in fewer than 3 dimensions have been extensively studied in the literature (see, e.g., \cite{lemos1,witten2}), although the observed astrophysical black holes are 3-dimensional.
Besides,
geometric structures in the form of a planar BH may have interesting applications in condensed matter physics (see, e.g., \cite{Hartnoll:2009sz}), in particular in the 2+1-dimensional superconductor \cite{bobev,albash,bilic4}.

Second, the effects of the curvature of the horizon, e.g., spherical or hyperbolic, are secondary in most situations of practical interest and could, effectively, at any rate be absorbed into a redefinition of the effective mass of the scalar perturbation.
Furthermore, experiments that simulate horizon-related phenomena, such as the Hawking effect, in analogue systems involving the flow of water in a basin, Bose-Einstein condensates, or cold atoms in a trap, often employ a linear setup, making planar black holes a more suitable choice for practical purposes. \cite{Hossenfelder:2017iom,Bilic:2022psx}.
In ultrarelativistic heavy-ion collisions, the fluid of particles is predominantly produced along one space dimension. Hence, the effective spactime is 1+1, which is equivalent to planar geometry where two space dimensions may be ignored.

Third, we provide a \emph{proof of concept} on how analogue Lagrangians for a general class of space-times, not just with individual metrics, can be constructed, thereby significantly extending the menagerie of known analog black hole metrics. The case of generic black hole space-times, not only for specific blackening factors, provides a suitable starting point due to its relevance for phenomena involving an event horizon, which is one of the main research points in analog gravity experiments, and due to previous work on which to build.

The paper is organized as follows. 
In section \ref{sec:outline} we define the geometry and its conformally rescaled metric. 
In  section \ref{sec:lagrangian} we outline a field theory description of a fluid
 and derive the propagation equation for acoustic perturbations. 
The main result follows in section \ref{relativistic}, where we show how a generic planar black hole metric can be mapped to the effective geometry of a fluid in which acoustic perturbations propagate. 
 In section \ref{entangle} we define the analog entanglement entropy for a general analog planar BH metric and compute it numerically for an analog planar AdS$_5$ BH.
	Concluding remarks are given in section \ref{sec:conclusions}.

We adopt a convention in which  the speed of light  and Planck constant $\hbar$ are set to unity, 
$c$  denotes  the speed of sound, and the metric signature is `mostly plus', i.e., $\{-,+,\ldots,+\}$.

\section{Conformal rescaling}
\label{sec:outline}

For the purpose of being self-contained, we summarize a result from \cite{Hossenfelder:2017iom}, which
shows how an additional degree of freedom in the form of a conformal factor can be introduced into an analog metric.

Consider a space-time in $n+1$ dimensions conformal to a rather generic stationary planar 
BH metric, which, for later convenience, we take to be parameterized as
\begin{eqnarray}
ds^2 = G_{\mu\nu}dx^\mu dx^\nu=\frac{\Omega(t,x,z)^2}{ \sqrt{1-\gamma(z)} } 
\left[ -\gamma(z) dt^2 + \frac{dz^2}{\gamma(z)} + d \mbox{\boldmath $x$}^2 \right] , 
\label{eq:metric}
\end{eqnarray}
where the function $\gamma$ is referred to as the  `blackening factor'.
The metric as written in (\ref{eq:metric}) refers to a general planar metric. In the next section, we will show how this form of metric can be achieved as an effective
acoustic metric with the help of a specifically designed fluid flow. 
If there is a horizon located at $z=\ell$, where $\gamma(\ell)=0$, the outside region is characterized by $\gamma > 0$. A canonical scalar field ${\varphi}$ propagating in this background with effective mass $m_{\rm eff}$ 
satisfies the equation of motion \cite{birrell_davies_1982}
\begin{eqnarray}
\square \varphi -  m_{\rm eff}^2\varphi\equiv
\frac{1}{\sqrt{|G|}} \partial_\mu \left( \sqrt{|G|}\, G^{\mu\nu} \partial_\nu \varphi \right) 
-m_{\rm eff}^2\varphi=0,
\label{eq:scalar_eom}
\end{eqnarray}
where it is assumed that $m_{\rm eff}$ and $\varphi$ depend on the coordinates $t$, $x$, $y$, and $z$.
The symbol $\square$ denotes the Klein-Gordon operator in curved space with the metrics $G_{\mu\nu}$.
Via a rescaling\footnote{Note that we use a slightly different convention than in \cite{Hossenfelder:2017iom}.} $\varphi = \Omega^{\frac{1-n}{2}}\wt{\varphi}$, this equation is equivalent to the conformally rescaled equation of motion \cite{birrell_davies_1982,Hossenfelder:2017iom}
\begin{eqnarray}
\tilde{\square} \tilde{\varphi} - \wt{m}_{\rm eff}^2(t,z,x) \wt{\varphi} = 0,
\label{eq:scalar_eom_new}
\end{eqnarray}
where
$\tilde{\square}$ denotes the Klein-Gordon operator   in curved space with  the metrics $\tilde{G}_{\mu\nu}=\Omega^{-2} G_{\mu\nu}$.
The rescaled field $\wt{\varphi}$ is propagating in the background geometry with a conformally rescaled line element
\begin{eqnarray}
d\wt{s}^2 = \Omega(t,x,z)^{-2} ds^2
=\wt{G}_{\mu\nu}dx^\mu dx^\nu
= \frac{1}{ \sqrt{1-\gamma(z)} } \left[ -\gamma(z) dt^2 + 
\frac{dz^2}{\gamma(z)} + d \mbox{\boldmath $x$}^2 \right] ,
\label{eq:metric_new}
\end{eqnarray}
and effective mass squared  
\begin{eqnarray}
\wt{m}_{\rm eff}^2 = \Omega^2 m_{\rm eff}^2 + \Omega^{1/2 -n/2} \wt{\square} \Omega^{n/2-1/2} .
\label{eq:potential_new}
\end{eqnarray}

\section{The Lagrangian}
\label{sec:lagrangian}

We begin this section by introducing the Lagrangian formalism suitable for
description of generally nonisentropic fluids.
We mainly use notation and definitions from previous work (see, e.g., \cite{Bilic:2018fsk,Bilic:1999sq,Babichev:2007dw}), which are standard for this system.
Consider a Lagrangian as follows:
\begin{eqnarray}
{\cal L} = F(\chi) - V(\theta,t,x,y,z) ,
\label{eq:1agrangian}
\end{eqnarray}
where $\theta$ is a dimensionless scalar field.  The quantity  $F$ is an arbitrary function of the kinetic energy term
\begin{eqnarray}
\chi = - g^{\mu\nu} \theta_{,\mu} \theta_{,\nu} ,
\label{eq0102}
\end{eqnarray}
where $g^{\mu\nu}$ is the inverse metric of the background spacetime. 
We will shortly demonstrate that the Lagrangian (\ref{eq:1agrangian})  is associated with any perfect fluid 
given its equation of state. Besides, it has been shown that this Lagrangian, with the kinetic term $F$ only, in the so-called 
Thomas-Fermi  approximation corresponds to a canonical complex field Lagrangian that describes a  Bose-Einstein condensate
 (see, e.g.,  \cite{bilic2,bilic3}).

The energy-momentum tensor for \eqref{eq:1agrangian} is
\begin{eqnarray}
T_{\mu\nu} =  2 {\cal L}_\chi \theta_{,\mu}\theta_{,\nu}+ {\cal L} g_{\mu\nu},
\label{eq:se-tensor}
\end{eqnarray}
where the subscript $\chi$ 
denotes a partial derivative with respect to $\chi$.
For $\chi>0$, this energy-momentum tensor  
will  describe a perfect fluid if we identify the pressure and energy density as
\begin{eqnarray}
p= {\cal L} ,
\label{eq02}
\end{eqnarray}
\begin{eqnarray}
\rho = 2\chi {\cal L}_{\chi} - {\cal L},
\label{eq03}
\end{eqnarray}
and the fluid velocity vector as 
\begin{eqnarray}
u_{\mu} = \frac{\theta_{,\mu}}{ \sqrt{\chi} }.
\label{eq:potential_flow_equation}
\end{eqnarray}
This equation describes the so-called `potential flow'.  
Solutions of this form are the relativistic analogue
of potential flow in non-relativistic fluid dynamics \cite{landau1959fluid}
and is usually ascribed to isentropic and irrotational flows.
Isentropic flow is characterized by the vanishing of the gradient 
$s_{,\mu}=0$, with $s$ being the specific entropy, i.e., the entropy per particle.
In general, 
a flow may  be non-isentropic and have a non-vanishing
vorticity
$\omega_{\mu\nu}$ 
defined as
\begin{equation}
	\omega_{\mu\nu}= h^{\rho}_{\mu}
	h^{\sigma}_{\nu} u_{[\rho;\sigma]},
	\label{eq203}
\end{equation}
where
\begin{equation}
	h^{\mu}_{\nu} =
	\delta^{\mu}_{\nu}+
	u^{\mu}u_{\nu} .
	\label{eq303}
\end{equation}
This tensor projects an arbitrary vector in space-time
into its component in the subspace orthogonal to
$u^{\mu}$.
If the conditions of isentropy and  vanishing vorticity are  assumed, the velocity field may be expressed by 
\begin{equation}
	w u_\mu =\theta_{,\mu}
	\label{eq403}
\end{equation}
where $\theta$ is the velocity potential and $w$ is the specific enthalpy. 
The reverse of the above statement is not true: a potential flow alone implies
only vanishing vorticity and 
implies neither isentropy nor particle number conservation.
In a potential flow, as may be easily shown \cite{Bilic:2018fsk}, the entropy gradient
is proportional to the gradient of the potential, i.e.,
\begin{equation}\label{rnif10}
	s_{,\mu}=w^{-1} u^{\nu}s_{,\nu}\theta_{,\mu} .
\end{equation}
The assumption (\ref{eq403}) is equivalent to (\ref{eq:potential_flow_equation}) if we identify
\begin{eqnarray}
w \equiv\frac{p+\rho}{n} = \sqrt{\chi}.
\label{eq:enthalpy}
\end{eqnarray}
Hence, the potential flow is automatically satisfied in the field-theory formalism
with a scalar field $\theta$ playing the role of the velocity potential.
Furthermore, in view of (\ref{eq:enthalpy}) with (\ref{eq02}) and (\ref{eq03}), we identify the particle number density as
\begin{eqnarray}
n  = 2\sqrt{\chi}{\cal L}_\chi .
\label{eq:particle_density} 
\end{eqnarray}
This is consistent with the Gibbs relation
\begin{eqnarray}
dp =  n dw - n T ds
= {\cal L}_{\chi} d \chi + {\cal L}_{\theta} d \theta ,
\label{eq:Gibbs_relation}
\end{eqnarray}
when a functional relationship $s = s(\theta)$ is assumed.

Thus, we have constructed a field theory description of a fluid.
Following \cite{Bilic:2018fsk}, the ideal irrotational fluid will satisfy the
Euler equation – i.e., the energy momentum conservation – if,
in addition to the potential flow equation \eqref{eq:potential_flow_equation}, the field satisfies the equation of motion
\begin{equation}
	(2{\cal L}_\chi g^{\mu\nu}\theta_{,\nu})_{;\mu}+\frac{\partial{\cal L}}{\partial \theta}=0 .
	\label{eq6}
\end{equation}
Using \eqref{eq:potential_flow_equation} and \eqref{eq:particle_density}), this equation  can  be written  as
\begin{eqnarray}
(n u^{\mu})_{;\mu} &=& \frac{\partial {V}}{\partial \theta}.
\label{eq:bg_eom}
\end{eqnarray}

Next, we briefly describe the derivation of the propagation equation for linear perturbations
of a nonisentropic flow assuming a fixed background geometry.
Given  some average bulk motion represented by
$p$, $n$, and
$u^{\mu}$, following the standard procedure \cite{Visser:1997ux} we make a replacement
\begin{equation}
	p\rightarrow p+\delta p, \quad n\rightarrow n+\delta n ,
	\quad
	u^{\mu}\rightarrow u^{\mu}+\delta u^{\mu},
	\label{eq008}
\end{equation}
where the perturbations
$\delta p$,
$\delta n$, and
$\delta u^{\mu}$
are induced by
a small perturbation $\theta = \theta_0 + \delta\theta$,
around the background $\theta_0$.
From
(\ref{eq403}) we find
\begin{equation}
	\delta w=-u^\mu\delta\theta_{,\mu},
	\label{eq406}
\end{equation}
\begin{equation}
	w\delta u^\mu=(g^{\mu\nu}+u^\mu u^\nu)\delta\theta_{,\nu}.
	\label{eq407}
\end{equation}
Using this and (\ref{eq008}) equation (\ref{eq:bg_eom}) at linear order yields
\begin{equation}
	\left(f^{\mu\nu}
	\delta\theta_{,\nu} \right)_{;\mu}
	+\left[\left(
	\frac{\partial n}{\partial\theta}u^\mu\right)_{;\mu}
	-\left(\frac{\partial^2 V}{\partial\theta^2}\right)\right]\delta\theta=0,
	\label{eq413}
\end{equation}
where
\begin{equation}
	f^{\mu\nu}=\frac{n}{w}\left[g^{\mu\nu}+\left(1-\frac{w}{n}\frac{\partial n}{\partial w}\right)u^\mu u^\nu\right] .
	\label{eq423}
\end{equation}
Then, it may be easily shown
that equation (\ref{eq413}) can be recast into the form
\begin{eqnarray}
|\wt{G}|^{-1/2} ( |\wt{G}|^{1/2}\wt{G}^{\mu\nu}  \delta\theta_{,\nu} )_{;\mu}
-m_{\rm eff}^2 \delta\theta=0  ,
\label{eq:perturbation_eom}
\end{eqnarray}
where $\wt{G}^{\mu\nu}$ is the inverse  of
the relativistic acoustic metric \cite{Bilic:1999sq}
\begin{eqnarray}
\wt{G}_{\mu\nu} = \frac{n}{m^2 w c} \left[ g_{\mu\nu} + (1-c^2) u_\mu u_\nu \right]
\label{eq:rel_ac_metric}
\end{eqnarray}
with determinant $\wt{G}$.
 Hence, the acoustic perturbation $\delta\theta$ is a scalar field equivalent to the field $\wt{\varphi}$ in section \ref{sec:outline},  that satisfies the Klein-Gordon equation (\ref{eq:scalar_eom_new}) equivalent to (\ref{eq:perturbation_eom}).
Note that in this section, unlike in the section \ref{sec:outline},  the metric $\wt{G}_{\mu\nu}$ with capital $G$ denotes the effective acoustic metric, whereas the metric of the background spacetime is denoted by $g_{\mu\nu}$.
The quantity $\tilde{m}_{\rm eff}$ in Eq.\ (\ref{eq:perturbation_eom}) and henceforth 
is assumed to depend on $t$, $x$, $y$, and $z$. 
An arbitrary mass parameter $m$ in (\ref{eq:rel_ac_metric}) is introduced to make the metric in $\wt{G}_{\mu\nu}$ dimensionless, and $c$ is the speed of sound defined by
\begin{eqnarray}
\left. c^2 \equiv \frac{\partial p}{\partial \rho} \right|_s
= \left.\frac{n}{w}\frac{\partial w}{\partial n} \right|_s .
\label{eq04} 
\end{eqnarray}
	From now on, we will assume that the sound speed satisfies
	\begin{equation}
		0\leq c \leq 1.
		\label{eq001}
	\end{equation}
The quantity $m_{\rm eff}$ is the effective mass  defined by
\begin{eqnarray}
m^2\sqrt{|\wt{G}|}\,m_{\rm eff}^2 =
\left.  \frac{\partial^2 V}{\partial \theta^2}\right|_{\theta_0}  .
\label{eq:m_eff}
\end{eqnarray}
Using (\ref{eq:1agrangian})-(\ref{eq:potential_flow_equation}) one can derive
the relation
\begin{eqnarray}
m^2\sqrt{|\wt{G}|}\,\wt{G}^{\mu\nu} =
\left.- \frac{\partial^2 F}{\partial \theta_{,\nu} \partial\theta_{,\mu} }\right|_{\theta_0}  ,
\label{eq:pert_metric}
\end{eqnarray}
also  derived by Babichev et al.\ \cite{Babichev:2007dw} in a different context.

Here, it is worth mentioning the diffeomorphism invariance of our analog model.
The Lagrangian $\mathcal{L}$ defined in (\ref{eq:1agrangian}) with (\ref{eq0102}) is a scalar, so the action $S=\int d^4x \sqrt{-g} \mathcal{L}$ and the corresponding field equations are invariant under general coordinate transformation. The perfect fluid stress tensor $T_{\mu\nu}$  defined above with scalar variables $\rho$ and $p$, and the four-vector $u^\mu$ is a covariant tensor.
Furthermore, the manipulations leading to the acoustic metric (\ref{eq:rel_ac_metric}) are fully covariant.
Hence, the hydrodynamic model that stems from the on-shell Lagrangian and the derived acoustic geometry are 4D-diffeomorphism invariant.

\section{Relativistic acoustic metric}
\label{relativistic}

Building on \cite{Bilic:2018fsk}, we now proceed to show that an acoustic perturbation in a 
fluid -- the dynamics of which is described by an explicit field theory Lagrangian -- can be realized 
as a scalar field propagating in the background \eqref{eq:metric_new}.
This extends the procedure developed in \cite{Bilic:2018fsk}, which was restricted  
to a  planar AdS$_5$ BH with 
\begin{equation}
	\gamma(z)=1-\frac{z^4}{\ell^4},
	\label{eq01}
\end{equation}
where the function   $\gamma (z)$ is the blackening coefficient in the planar AdS$_5$ metric with the horizon at $z=\ell$, similar to the Schwarzschild metric where $\gamma$ depends only on $r$ with the horizon at 
$r=r_{\rm Sch}$.
Here we will show 
that the formalism can be generalized to simulate metrics of the form \eqref{eq:metric_new} 
with an arbitrary blackening factor $\gamma(z)$ 
	subject only to the restriction 
	\begin{equation}
		 \gamma \leq 1.
		\label{eq002}
	\end{equation}
In particular, we will show that an acoustic perturbation propagating in a fluid described by the Lagrangian 
of the form \eqref{eq:1agrangian} represents an analogue dual of a scalar field 
propagating in the background \eqref{eq:metric_new}. 
In other words, if a fluid is described by the Lagrangian \eqref{eq:1agrangian}, the dynamics of acoustic perturbations 
described by (\ref{eq:perturbation_eom}-\ref{eq:m_eff})
will have the form of the Klein-Gordon equation \eqref{eq:scalar_eom_new} 
in a curved space-time described by the line element \eqref{eq:metric_new}.

The first step is to bring the metric  \eqref{eq:metric_new}  to a form that can be compared 
with the acoustic metric (\ref{eq:rel_ac_metric}).
For this purpose, we make the following coordinate transformation from the coordinates $t$ and $z$ to new coordinates $\tilde{t}$ and $\tilde{z}$, keeping $x$ and $y$ intact,
\begin{eqnarray}
t =\wt{t} + f(z),\quad z = g(\wt{z}).
\label{eq0101}
\end{eqnarray}
Then, the line element from \eqref{eq:metric_new} takes the form
\begin{eqnarray}
d\wt{s}^2 &=& \frac{1}{ \sqrt{1-\gamma}} \biggl\{ - d\wt{t}^2 + d\wt{z}^2 + d \mbox{\boldmath $x$}^2 \nonumber\\
& & + \left[ (1-\gamma) d\wt{t}^2 - 2 \sqrt{(1-\gamma)(c^2-\gamma)} 
d\wt{t} d\wt{z} + (c^2-\gamma) d\wt{z}^2 \right] \biggr\},
\label{eq:metric_II}
\end{eqnarray}
where
\begin{eqnarray}
\frac{dg}{d\wt{z}}  =  c , \quad 
\frac{df}{d z} = \frac{\sqrt{(1-\gamma)(c^2-\gamma)}}{c \gamma} .
\end{eqnarray}
Comparing with \eqref{eq:rel_ac_metric} allows one to read off the non-vanishing components of the 4-velocity
\begin{eqnarray}
u_{\wt{t}} = \sqrt{\frac{1-\gamma}{1-c^2}} , \quad
u_{\wt{z}} = -\sqrt{\frac{c^2-\gamma}{1-c^2}} .
\label{eq9}
\end{eqnarray}
	These equations imply
\begin{equation}
	\gamma\leq c^2 \leq 1 .
	\label{eq151}
\end{equation}

Next, assuming a potential flow (\ref{eq403})
we derive closed expressions for $w$, $n$, and $c$ in terms of the variable $\wt{z}$. 
Since the metric is stationary, the velocity potential must be of the form
\begin{equation}
	\theta=m\wt{t} + h(z) ,
	\label{eq:bg_parametrization}
\end{equation}
where $m$ is an arbitrary mass and $h(z)$ is a function of $\wt{z}$ through $z=g(\wt{z})$.
The specific enthalpy is then given by
\begin{eqnarray}
w = \frac{m}{u_{\wt{t}}}
= m \sqrt{\frac{1-c^2}{1-\gamma}} 
\label{eq:w_identification}
\end{eqnarray}
and the function $h(z)$ is determined through 
\begin{eqnarray}
\frac{dh}{dz} 
= -\frac{m}{c} \sqrt{\frac{c^2-\gamma}{1-\gamma}} .
\end{eqnarray}
From the definition (\ref{eq04}) 
it follows 
\begin{eqnarray}
c^2	
=	\frac{n}{w} \frac{\partial_{\wt{z}}w}{\partial_{\wt{z}}n}.
\label{eq05}	
\end{eqnarray}
This implies that the sound speed $c$ must satisfy 
a  differential equation
\begin{eqnarray}
\frac{\partial}{\partial \wt{z}} c^2
=\left( c^2 - \frac{1}{2} \right) \frac{\partial}{\partial \wt{z}} \ln (1-\gamma) ,
\end{eqnarray}
with solution
\begin{eqnarray}
c^2	=	c_1 (1-\gamma) + \frac{1}{2}.
\label{eq10}
\end{eqnarray}
Due to the requirement (\ref{eq001}), the integration constant $c_1$ is restricted to 
\begin{equation}
-\frac{1}{2(1-\gamma_{\rm min})} \leq c_1 \leq \frac{1}{2(1-\gamma_{\rm min})} ,
\end{equation}
where $\gamma_{\rm min}$ is the minimal value of $\gamma$. If we do not want to cover
the region within the horizon, we can choose $\gamma_{\rm min}=0$, in which case 
we have 
$ -1/2 \leq c_1 \leq 1/2$.
Considering that the sound speed must  satisfy both (\ref{eq151}) and (\ref{eq10}), it is unlikely that with a single choice of $c_1$ we could cover the whole physical range $z>0$. We will elaborate more on this below.
Furthermore, in view of \eqref{eq:rel_ac_metric}, \eqref{eq:metric_II}, and 
(\ref{eq10}),
we can write the enthalpy and particle number density as 
\begin{eqnarray}
w = m \sqrt{\frac{1}{2(1-\gamma)} - c_1} ,  
\label{eq:wn_new}
\end{eqnarray}
\begin{eqnarray}
n =  m^3 \sqrt{\frac{1}{4(1-\gamma)^2} - c_1^2}= m^2w \sqrt{\frac{1}{2(1-\gamma)} + c_1}.
\label{eq8}
\end{eqnarray}
In principle, $c_1$ could be a function of $s$. However, since $w$ and $s$ are considered as independent variables, 
the right-hand side of \eqref{eq:wn_new} admits no explicit $s$-dependence. Hence, 
a consistent choice is $c_1\equiv \rm const$.
From \eqref{eq:wn_new} and \eqref{eq8} it follows
\begin{eqnarray}
n\frac{\partial w}{\partial \wt{z}}
= \frac{ m^2}{2} \sqrt{\frac{1}{2(1-\gamma)} + c_1} \:
\frac{\partial w^2}{\partial \wt{z}} 
= \frac{ m^4}{3} \frac{\partial}{\partial \wt{z}} \left(\frac{1}{2(1-\gamma)} + c_1 \right)^{3/2}.
\end{eqnarray}
Then, according to \eqref{eq:Gibbs_relation} the pressure reads
\begin{eqnarray}
p = \frac{ m^4}{3} \left(\frac{1}{2(1-\gamma)} + c_1 \right)^{3/2} - c_2(s) ,
\end{eqnarray}
where $c_2(s)$ is an arbitrary function of $s$.
In view of \eqref{eq:wn_new} the pressure  can also be expressed as
\begin{eqnarray}
p = \frac{ m^4}{3} \left(\frac{w^2}{m^2} + 2 c_1 \right)^{3/2} - c_2(s) .
\end{eqnarray}
This expression is precisely of the form \eqref{eq:1agrangian} in which 
\begin{eqnarray}
F(\chi)  =  \frac{ m^4}{3} \left(\frac{\chi}{m^2} + 2 c_1 \right)^{3/2},
\label{eq15}
\end{eqnarray}
$c_2$ is identified with $V$ and the specific enthalpy with $\sqrt{\chi}$ as in \eqref{eq:enthalpy}.  

Therefore, we have shown that the Lagrangian \eqref{eq:1agrangian} with \eqref{eq15} can be used to construct 
an analogue model for a scalar field propagating in the metric \eqref{eq:metric_new} with an arbitrary $\gamma(z)$.
However, as mentioned previously, with a specific choice of constant $c_1$, we would, in general, only cover a part of the range   $z\geq 0$. 
If we require that the horizon $\gamma=0$ lies within the allowed range, we will find a constraint
as to how close to the limit $\gamma=1$ our analog metric is applicable. 
Suppose we choose to cover only the outside region, so that $c_1$ is restricted to the interval [-1/2,1/2]. As a consequence of Eq.\  (\ref{eq151}), our analog model will break down at 
a point $z=z_{\rm min}$ which is
the maximal root of the algebraic equation $\gamma(z)=2/3$.
This equation follows from imposing
$c^2=\gamma$ and Eq. (\ref{eq10}) with maximal $c_1=1/2$.
For example, for a planar AdS$_5$ BH with $\gamma=1-z^4/\ell^4$ one finds
$z_{\rm min}=\ell/3^{1/4}$. Similarly, if we chose to cover the entire region within the horizon up to $z=\infty$, the algebraic equation would read $\gamma(z)=1/2$. Then, for 
a planar AdS$_5$ we would obtain $z_{\rm min}=\ell/2^{1/4}$.

It is worth noting that the Lagrangian (\ref{eq:1agrangian}) with (\ref{eq15}) has the same functional dependence on $\chi$ as the one 
found in \cite{Bilic:2018fsk}, where it was derived from the requirement that the analogue metric  correctly 
reproduces the planar AdS$_5$ BH.  Hence, the functional form (\ref{eq15}) is generic.
However, the fluid dynamics is not completely determined unless the potential $V(\theta)$ is specified
because the flow velocity components are fully determined by the velocity potential $\theta$, which solves the field equations. 
To find a solution to the field equations, we need to
specify the potential $V(\theta)$ which  
will be done in the following section.

\subsection{The potential}

Recall that we are considering a scalar field $\theta=\theta_0 + \delta\theta$, i.e., a small acoustic perturbation $\delta\theta$ around a fixed background $\theta_0$.
The equation of motion of this perturbation is an analog model of a particle propagating in the curved space-time. Then, the potential $V$ has to meet a requirement that its first derivative, when evaluated on the background \eqref{eq:bg_parametrization}, is determined by the equation \eqref{eq:bg_eom}.
In applications where one wishes to simulate a specific effective mass\footnote{Such as, e.g., in \cite{Hossenfelder:2018lhy}} in addition to a specific metric, equation \eqref{eq:m_eff} requires imposing conditions on the second derivative of $V$.
Thus, the potential $V$ has to be chosen such that
\begin{eqnarray}
\left.	\frac{\partial V}{\partial \theta} \right|_{\theta=\theta_0}
	= (n u^{\mu})_{;\mu};,  \qquad
\left.	\frac{\partial^2 V}{\partial \theta^2} \right|_{\theta=\theta_0}
	= \sqrt{|\wt{G}|} m^2 m_{\rm eff}(\wt{z}) ,
	\label{eq:potential_cond}
\end{eqnarray}
where the new coordinates $\wt{t}$ and  $\wt{z}$ are defined by the coordinate transformation (\ref{eq0101}).
In principle, one could satisfy these conditions in many ways.
Quite generally, a suitable  potential can be written as
\begin{eqnarray}
	V= \alpha(\wt{z}) \theta f_1(\theta/\theta_0)+\beta(\wt{z})\theta^2 f_2(\theta/\theta_0)
	\label{eq20}
\end{eqnarray}
where $f_1(x)$ and $f_2(x)$ are arbitrary functions which at $x=1$ (i.e., when $\theta=\theta_0$) satisfy
\begin{eqnarray}
\left. \left(xf_1(x) \right)''\right|_{x=1}=0 \;,  \qquad
\left. \left(x^2f_2(x) \right)'\right|_{x=1}=0	\, .
\label{21}
\end{eqnarray}
and $\alpha(\wt{z}), \beta(\wt{z})$ are chosen to match \eqref{eq:potential_cond}.
Therefore, the potential $V$ will generally have
to be chosen coordinate-dependent.
This would present no real obstacle from a practical point of view, as experimental setups for analog gravity with moving and oscillating horizons are already being conducted (see e.g., Refs.\ \cite{Steinhauer:2014dra,Steinhauer:2015saa}), and time- and position-dependent external potentials could be simulated with the same setup.

From a theoretical point of view, there could be some caveat that limits the choice of potentials. 
That comes from the condition that the Gibbs relation \eqref{eq:Gibbs_relation} must hold. 
At first sight, it may seem a bit odd how the relation containing  only two degrees of freedom could be satisfied 
with a generic potential $V(\theta,t,z,\mbox{\boldmath $x$})$. 
However, one has to keep in mind that the functional identities in section \ref{sec:lagrangian} 
are independent of the specific coordinate dependence of the potential, and the crucial point 
is that the Gibbs relation \eqref{eq:Gibbs_relation} has to hold as an on-shell functional identity. 
This is to say that it must be possible to express the pressure $p$ as a functional depending on two variables $w$ and $s$ 
which are defined on the function space of solutions to the equations of motion.
This reduces the effective number of degrees of freedom.\footnote{Note that the Gibbs relation need 
	not hold for a generic field that does not satisfy the equations of motion.}
In practice, however, it could be rather non-trivial to check \eqref{eq:Gibbs_relation} explicitly, 
and the following construction might be more convenient.

Assume a Lagrangian with no explicit coordinate dependence of the form
\begin{eqnarray}
	{\cal L} = F(\chi) - V(\theta) .
	\label{eq:1agrangian_simple}
\end{eqnarray}
The Gibbs relation \eqref{eq:Gibbs_relation} is now automatically satisfied and for 
a solution $\theta_0$ of the equations of motion \eqref{eq:bg_eom}, the analogue metric and effective mass for a perturbation follow from (\ref{eq:pert_metric},\ref{eq:m_eff}). 
In this situation, one can 
proceed to construct a potential $V$ that reproduces the desired analog 
metric in a way similar to \cite{Bilic:2018fsk}, where it has been worked out 
for the case of a planar BH in AdS space-time.
In order to then explicitly match the effective mass to a desired value, consider the Lagrangian \eqref{eq:1agrangian_simple} changed by an ${\cal O}(\theta-\theta_0)^2$ deformation around the found background solution $\theta_0$, e.g.
\begin{eqnarray}
	{\cal L}'  =  F(\chi) - V(\theta) - \frac{a(\theta,t,x,z)}{2} (\theta-\theta_0)^2
	\label{eq:1agrangian_simple_def}
\end{eqnarray}
By construction, $\theta_0$ is still a solution to the equations of motion and all identities from section \ref{sec:lagrangian} will hold identically when evaluated for $\theta_0$, with the exception of \eqref{eq:m_eff}, which, as the only quantity in the perturbation equations, depends on second order derivatives of the Lagrangian with respect to $\theta$. Thus, the effective mass changes to
\begin{eqnarray}
	(m'_{\rm eff})^2
	&=& m_{\rm eff}^2 + \frac{a}{\sqrt{|\wt{G}|}} .
	\label{eq:m_eff_prime}
\end{eqnarray}
Therefore, by a suitable choice of $a(\theta,t,x,z)$, any $m'_{\rm eff}$ can be reproduced without changing the analog metric. 

Of course,  equation \eqref{eq:1agrangian_simple} has to remain an analog model when considering deviations around $\theta_0$, including the Gibbs relation \eqref{eq:Gibbs_relation}, which is the most crucial for the analog gravity construction to work. This, however, follows directly from the theorem of implicit functions, if  $\theta_0$ is not a degenerate point in the space of solutions.

\section{Analog entanglement entropy}
\label{entangle}

The entanglement entropy in general is defined for
a quantum system divided into two subsystems $A$ and $B$.
For the density of states
matrix $\rho=|\Psi\rangle \langle \Psi|$, we define the reduced density matrix for the subsystem $A$
by taking a partial trace over the subsystem $B$, i.e.,
$\rho_{A}=\mathrm{tr}_{B}\, |\Psi\rangle \langle \Psi|$.
Then, the entanglement entropy is defined as
\begin{eqnarray}
	S_A = - \mathrm{tr}_{A}\, (\rho_{A} \log \rho_{A}).
	\label{eq0}
\end{eqnarray}
The quantity $S_A$
is the entropy for an observer who can access information only from the subsystem $A$ and can receive no information from
$B$. The subsystem $B$ is analogous to the interior of
a BH horizon for an observer outside of
the horizon. However, it is often not easy to compute the entanglement entropy, in particular in field theory in
3+1 or higher dimensions.

As discussed previously, the prescription  for our analog model is only valid from the point $z_{\rm min}$
up to the horizon location at $z=\ell$.
Hence, we place the boundary of our model spacetime at $z_{\rm min}$ and cut off
the section  from $z=0$ to $z_{\rm min}$ as it has been done for AdS$5$ in the Randall-Sundrum model
\cite{randall1,randall2}. The plane at $z=z_{\rm min}$ defines the boundary of our analog spacetime, similar to the boundary of AdS spacetime at $z=0$.
Thus, our system is divided in two subsystems, $A$ and $B$, whee $A$ extends from
$z_{\rm min}$ up to the BH horizon at $z=\ell$ and $B$ from
$z=0$ to $z=z_{\rm min}$. Hence, the concept of entanglement entropy arises naturally in our
analog model.

A convenient description of the entanglement entropy is derived in an $n+1$-dimensional field theory. It has been shown that the leading term of the entanglement entropy 
can be expressed as the area law \cite{bombelli,srednicki}
\begin{equation}
	S_A=a \frac{\mbox{Area}(\partial
		A)}{\ell^{n-1}}+\mbox{subleading terms},
	\label{divarea}
\end{equation}
where $\partial A$ is the boundary of $A$, $\ell$ is an ultraviolet cutoff or the minimal length in the theory, and  $a$ is a constant which depends on the system. 
It is not accidental that this area law is of the same form as  the Bekenstein-Hawking entropy of BHs in 3+1 dimensions, which is proportional to the area of the event horizon, 
with the constants $n=3$, $a=1/4$, and $\ell$ equal to the Planck length. 

As we are dealing with an analog geometry, we will assume the existence of a minimal length. This length is typically of the order of the atomic separation. Below this scale, the bulk description of the fluid fails. 
This length describes the distance over which the wave function of a BE condensate tends to its bulk value when subjected to a localized perturbation. It is referred to as the healing length \cite{pethick}.
In analog gravity systems,
a healing length $\ell_{\rm hl}$
plays the role of the Planck length \cite{uhlmann,girelli,fleurov,rinaldi,anderson}
and for a BE gas is typically of order $\ell_{\rm hl} \simeq 1/(mc)$, 
where $m$ is the boson mass and $c$ is the sound speed. 

The entropy-area relation arises in the context of AdS/CFT duality. 
According to AdS/CFT, the entanglement entropy, being basically tied to the gravity in the bulk, should reflect fundamental features of the boundary gauge theory.
In this regard, we will study the so-called 
`holographic entaglement entropy' in 3+1 dimensions in the analog gravity context. 
In contrast to the usual entanglement entropy, for holographic entanglement entropy, the area of a fixed two-dimensional subsystem on the boundary depends on the geometry in the bulk.
We expect that the holographic entanglement entropy in the analogue model
discussed in section \ref{relativistic} should exhibit the features of the analog planar BH horizon.

The holographic entanglement entropy $S$ in a 2+1-dimensional boundary field theory is defined 
for a 2-dimensional subsystem $\Sigma$ that has an
arbitrary one-dimensional boundary $\partial\Sigma$. 
To calculate the entanglement entropy in our analog system, we use the area law prescription \cite{ryu,ryu2} 
\begin{equation}
	S=\frac{{\rm Area}(\Sigma)}{4\ell^2} .
	\label{eq6000}
\end{equation}
Here, $\Sigma$ is the two-dimensional static minimal-area surface in
the 3+1-dimensional bulk with boundary $\partial\Sigma$
and the scale $\ell$ we will identify with $\ell_{\rm hl}$.
\begin{figure}[t]
	\begin{center}
		\includegraphics[width=0.8\textwidth,trim= 0 0cm 0 0cm]{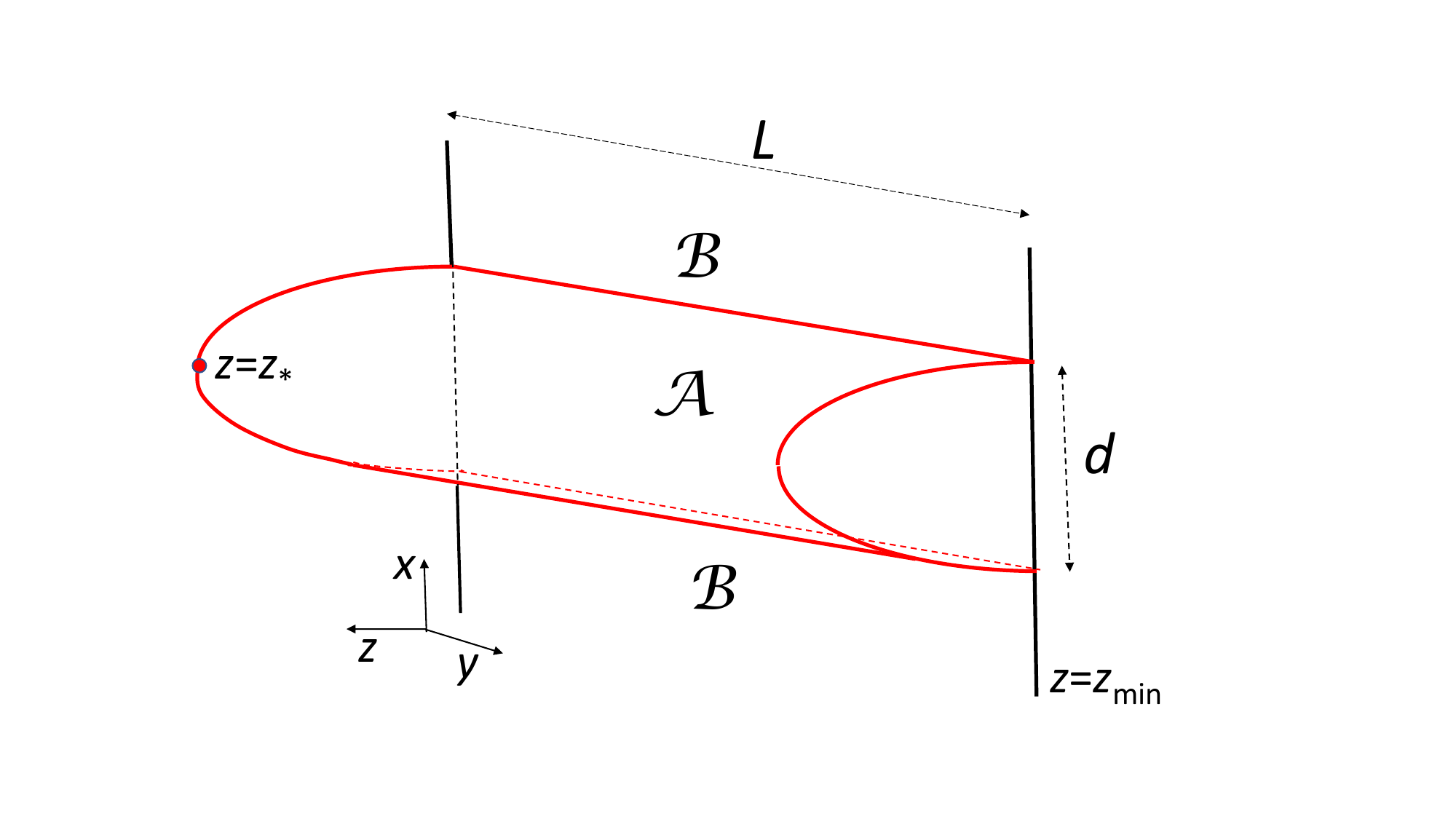}
		\caption{Strip geometry employed 
			to calculate the entanglement entropy. 
Adapted illustration from Ref.\ \cite{bilic4}.}
		\label{fig2}
	\end{center}
\end{figure}
We will apply the prescription (\ref{eq6000}) to the strip geometry suggested in Ref.\ \cite{ryu} (see also \cite{Bilic:2022psx}) 
illustrated in Fig.\ \ref{fig2}, and calculate the entropy $S$ as a function of the strip width $d$. 

Consider the bulk metric (\ref{eq:metric_new}) with $n=3$ and
a surface $\Sigma$
defined by the equation
\begin{equation}
	z-z(x)=0 .
	\label{eq6001}
\end{equation}
Here, $z(x)$ is a function of $x$ such that $\Sigma$ extends into the bulk and is 
bounded by the perimeter of $\mathcal{A}$  as illustrated in Fig.\ \ref{fig2}.
The induced metric $\sigma_{ij}$ on $\Sigma$ defines   
the line element
\begin{equation}
	ds_\Sigma^2=\sigma_{ij}dx^idx^j= \frac{1}{\sqrt{1-\gamma(z)}}\left[dx^2\left(1+\frac{{z'}^2}{\gamma(z)}\right)+y^2\right].
	\label{eq6002}
\end{equation}
Finding the minimal area of $\Sigma$ is equivalent to maximizing the functional
\begin{equation}
	I[z,z']=-{\rm Area}(\Sigma)/L= -\int dxdy \sqrt{\det \sigma_{ij}}/L =\int_{-d/2}^{d/2}dx \mathcal{L}.
	\label{eq6003}
\end{equation}
Here, $L$ and $d$ are respectively the length and width of the strip, and 
\begin{equation}
	\mathcal{L}= -\frac{1}{\sqrt{1-\gamma}}\left(1+\frac{{z'}^2}{\gamma}\right)^{1/2}
	\label{eq00}
\end{equation}
The extremum condition $\delta I=0$ yields the equation of motion for $z$.
We will employ the fact that the equation of motion is satisfied if and only if the Hamiltonian is a constant of motion. 
Using the conjugate momentum 
\begin{equation}
	\pi=\frac{\partial\mathcal{L}}{\partial z'},
	\label{eq1}
\end{equation}
the Hamiltonian is defined as
\begin{equation}
	\mathcal{H}=\pi z' - \mathcal{L} = \frac{1}{\sqrt{1-\gamma}}\frac{1}{(1+{z'}^2/\gamma)^{1/2}}.
	\label{eq2}
\end{equation}
Since $z=z_*$ and $z'=0$ at the bottom of the surface, we obtain the equation
\begin{equation}
	\frac{1}{\sqrt{1-\gamma(z_*)}}= \frac{1}{\sqrt{1-\gamma(z)}}\frac{1}{(1+{z'}^2/\gamma(z))^{1/2}},
	\label{eq4}
\end{equation}
from which we can  express $z'$ as
\begin{equation}
	z'=\pm \frac{\sqrt{\gamma(z)(\gamma(z)-\gamma(z_*))}}{\sqrt{1-\gamma(z)}} . 
	\label{eq55}
\end{equation}
Inserting this into (\ref{eq6003}) 
and changing the integration variable from $x$ to $z$ with $dx=dz/z'$, we obtain 
the entanglement entropy 
expressed as an integral over $z$
\begin{equation}
	S =\frac{{\rm Area}}{4\ell^2}=\frac{L}{2\ell^2}\int_{z_{\rm min}}^{z_*}
	dz\frac{\sqrt{1-\gamma(z_*)}}{\sqrt{1-\gamma(z)}}
	\frac{1}{\sqrt{\gamma(z)(\gamma(z)-\gamma(z_*))}}.
	\label{eq06}
\end{equation}
The location of the bottom $z_*$  of the extremal surface is related to the strip width
\begin{equation}
	d=2 \int_{-d/2}^{d/2}dx=2\int_{z_{\rm min}}^{z_*}
	dz
	\frac{\sqrt{1-\gamma(z)}}{\sqrt{\gamma(z)(\gamma(z)-\gamma(z_*))}}.
	\label{eq6005}
\end{equation}
Given blackening metric function $\gamma$, equations (\ref{eq06}) and (\ref{eq6005})
define the entropy $S$ as a parametric function of the strip width $d$
with the parameter $z_*$ ranging from $z_{\rm min}$ to $\ell$.

\begin{figure}[h]
	\begin{center}
		\includegraphics[width=\textwidth,trim= 0 0cm 0 0cm]{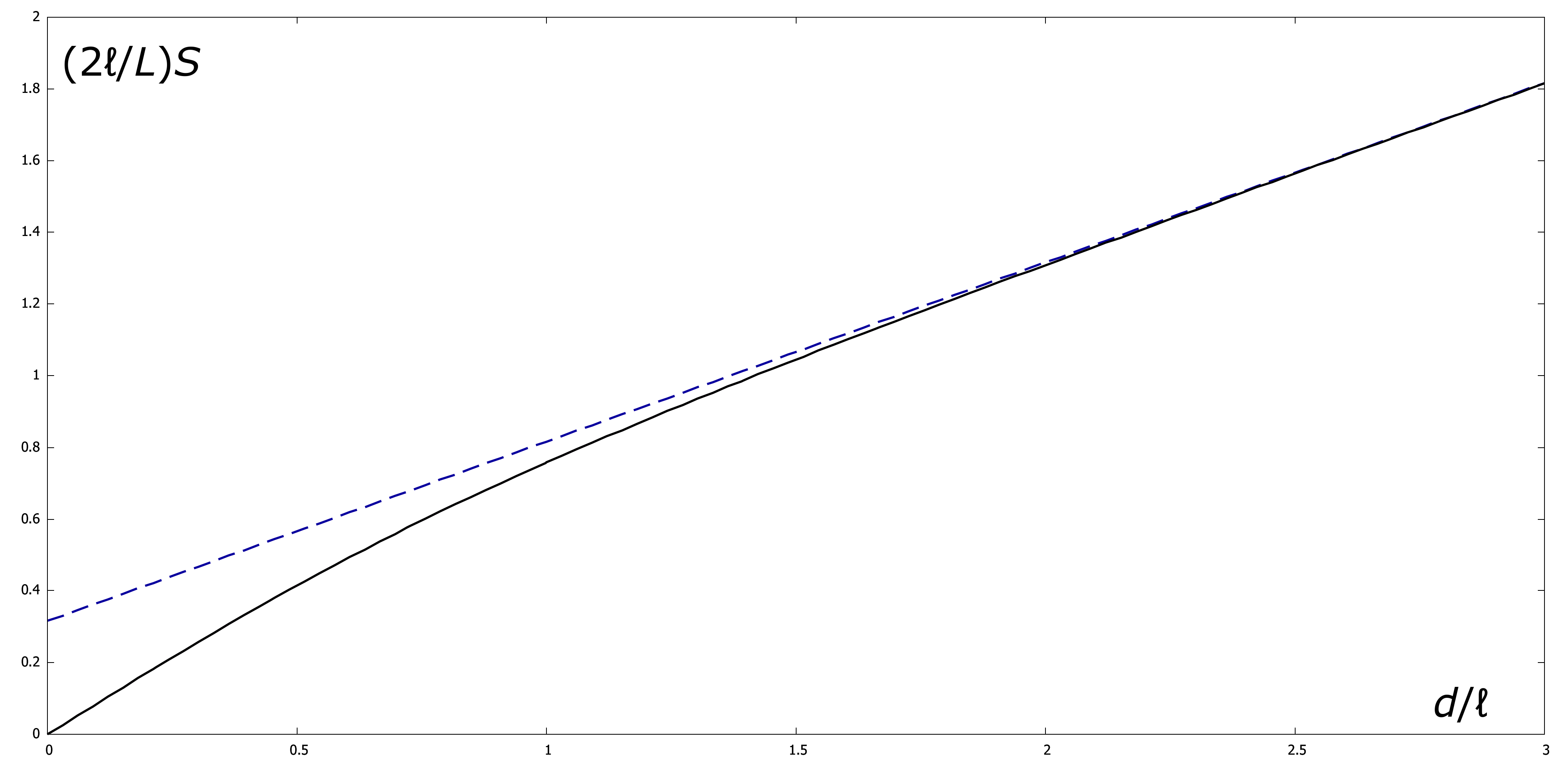}
		\caption{Holographic entanglement entropy (full black line) and limiting function 
			$S_{\rm lim}$ (blue dashed line) versus strip width.}
		\label{fig3}
	\end{center}
\end{figure}

By way of example, we numerically compute the function $S=S(d)$ for a planar AdS$_5$ BH with $\gamma$ as in Eq.\ (\ref{eq01}). Since the analog metric is 3+1-dimensional, we will ignore the fifth space coordinate, so the boundary at $z=z_{\rm min}$ will be 
a 2-dimensional space-like plane.
In the limit $z_* \rightarrow \ell$ both $S$ and $d$ diverge logarithmically. It may be easily shown that in this limit 
the function $S=S(d)$ asymptotically approaches the linear function
\begin{equation}
	S_{\rm lim}=\frac{L}{2\ell}\left(3^{1/4}-1+\frac{d}{2\ell}\right).
\end{equation}
Hence in the limit of large $d$, the entanglement entropy obeys the area law (\ref{divarea})
with $a=1/4$ and a subleading term equal to $(3^{1/4}-1)L/(2\ell)$.
In Fig.\ \ref{fig3}, we plot both functions $S(d)$ and $S_{\rm lim}(d))$ in units of
$L/(2\ell)$.

\section{Summary and conclusions}
\label{sec:conclusions}

Using the formalism of analogue gravity for the case of nonisentropic fluids from \cite{Bilic:2018fsk}, we have shown that by a suitable transformation of variables and choice of parametrization, a Lagrangian of the form \eqref{eq:1agrangian} is an analogue model for a scalar field propagating in a space-time that is conformal to a static, planar BH space-time. We have also demonstrated how, with a suitable adjustment of the external potential that couples to the analog Lagrangian, it is possible, for any given analog metric, to simulate an arbitrary effective mass for the perturbation.
	Furthermore, we have studied the analog entanglement entropy and computed it numerically for an analog planar AdS$_5$ BH.
These results are valid for a generic choice of conformal rescaling and blackening factor of the metric,  for an arbitrary effective mass of the scalar perturbation. 

It is worth noting that our acoustic metric is specified completely by the three independent functions: z-component of the velocity, the density $\rho$, and the pressure $p$ specified by the equation of state $p=p(\rho)$. Furthermore, the equation of continuity reduces these three degrees of freedom to two. Hence, as in a general acoustic geometry (see, e.g., Ref.\ \cite{Barcelo2011}), our analog geometry has two degrees of freedom per point in spacetime in contrast to the 3+1-dimensional pseudo-Riemannian geometry where the metric has 6 degrees of freedom.

The procedure outlined here allows for vastly extending the class of phenomena in gravity physics that can be simulated in condensed matter systems via the analogue gravity formalism.
Besides, the effects we have discussed may be of
phenomenological interest in all those phenomena that involve
relativistic fluids under extreme conditions.
For example, this may be the case in
ultrarelativistic heavy-ion collisions, where the fluid of particles is predominantly produced along one space dimension.

This class of phenomena has now been shown to include most non-rotating planar BH metrics considered in the literature -- as well as several cosmological space-times of particular interest. As our emphasis was put on planar BH geometries, our result also provides new foundations for the surge of investigations on how analog gravity interlinks with gauge/gravity duality\footnote{For reviews see e.g., \cite{Hartnoll:2009sz,Herzog:2011ec}} and condensed matter physics in the last years \cite{Das:2010mk,Semenoff:2012xu,Chen:2012uc,Khveshchenko:2013foa,Bilic:2014dda,Hossenfelder:2018lhy}, where the type of space-times considered here also plays a central role.

We emphasize again that the main difference between our analog model and analog planar models considered in the literature, e.g., in Refs. \cite{Bilic:2018fsk,Hossenfelder:2014gwa,Hossenfelder:2015pza,Dey:2016khw}, is in our study of a generic stationary planar BH metric. Besides, we provide a prescription for calculating the holographic entanglement entropy for a general analog planar BH spacetime with AdS asymptotic boundary.
A generalization to geometries with spherical or axial symmetry is possible and relatively straightforward, but will be left for future work.

\subsection*{Acknowledgement}
The work of N.B.\ has been supported by 
the ICTP-SEENET-MTP project NT-03 Cosmology-Classical and Quantum Challenges and the COST Action CA23130 --
BridgeQG: Bridging high and low energies in search of quantum gravity.
T.Z.\ acknowledges the support of the Swedish Research Council (Vetenskapsr\r{a}det) via grant No.\ 2015-04852.




\bibliography{APBHbib}

@book{landau1959fluid,
  title={Fluid mechanics},
  author={Landau, L.D. and Lifshits, E.M.},
  series={A-W series in advanced physics},
  url={https://books.google.se/books?id=v6kNAQAAIAAJ},
  year={1959},
  publisher={Pergamon Press}
}

@book{pethick,
  title={Bose-Einstein Condensation in Diluted Gases},
  author={Pethick, C.J. and Smith, H.},
  series={Condensed matter physics, nanoscience and mesoscopic physics},
  url={https://www.cambridge.org/hr/universitypress/subjects/physics/condensed-matter-physics-nanoscience-and-mesoscopic-physics/boseeinstein-condensation-dilute-gases-2nd-edition?format=HB&isbn=9780521846516},
  year={2008},
  publisher={Cambridge University Press}
}

@article{Babichev:2007dw,
      author         = "Babichev, Eugeny and Mukhanov, Viatcheslav and Vikman,
                        Alexander",
      title          = "{k-Essence, superluminal propagation, causality and emergent geometry}",
      journal        = "JHEP",
      volume         = "02",
      year           = "2008",
      pages          = "101",
      doi            = "10.1088/1126-6708/2008/02/101",
      eprint         = "0708.0561",
      archivePrefix  = "arXiv",
      primaryClass   = "hep-th",
      reportNumber   = "LMU-ASC-54-07",
      SLACcitation   = "%%CITATION = ARXIV:0708.0561;%%"
}

@Article{Barcelo2011,
author="Barcel{\'o}, Carlos
and Liberati, Stefano
and Visser, Matt",
title="Analogue Gravity",
journal="Living Reviews in Relativity",
year="2011",
month="May",
day="11",
volume="14",
number="1",
pages="3",
issn="1433-8351",
doi="10.12942/lrr-2011-3",
url="https://doi.org/10.12942/lrr-2011-3"
}

@article{Bilic:1999sq,
      author         = "Bili{\'c}, Neven",
      title          = "{Relativistic acoustic geometry}",
      journal        = "Class. Quant. Grav.",
      volume         = "16",
      year           = "1999",
      pages          = "3953-3964",
      doi            = "10.1088/0264-9381/16/12/312",
      eprint         = "gr-qc/9908002",
      archivePrefix  = "arXiv",
      primaryClass   = "gr-qc",
      SLACcitation   = "%%CITATION = GR-QC/9908002;%%"
}

@article{Bilic:2014dda,
      author         = "Bili{\'c}, Neven and Domazet, Silvije and Toli{\'c}, Dijana",
      title          = "{Analog geometry in an expanding fluid from AdS/CFT perspective}",
      journal        = "Phys. Lett.",
      volume         = "B743",
      year           = "2015",
      pages          = "340-346",
      doi            = "10.1016/j.physletb.2015.03.009",
      eprint         = "1410.0263",
      archivePrefix  = "arXiv",
      primaryClass   = "hep-th",
      SLACcitation   = "%%CITATION = ARXIV:1410.0263;%%"
}

@article{Bilic:2018fsk,
      author         = "Bili{\'c}, Neven and Nikoli{\'c}, Hrvoje",
      title          = "{Analog gravity in nonisentropic fluids}",
      journal        = "Class. Quant. Grav.",
      volume         = "35",
      year           = "2018",
      number         = "13",
      pages          = "135008",
      doi            = "10.1088/1361-6382/aac5c6",
      eprint         = "1802.03267",
      archivePrefix  = "arXiv",
      primaryClass   = "gr-qc",
      SLACcitation   = "%%CITATION = ARXIV:1802.03267;%%"
}

@book{birrell_davies_1982,
      place          = {Cambridge},
      series         = {Cambridge Monographs on Mathematical Physics},
      title          = {Quantum Fields in Curved Space},
      DOI            = {10.1017/CBO9780511622632},
      publisher      = {Cambridge University Press},
      author         = {Birrell, N. D. and Davies, P. C. W.},
      year           = {1982},
      collection     = {Cambridge Monographs on Mathematical Physics}
}

@article{Chen:2012uc,
      author         = "Chen, Pisin and Rosu, Haret",
      title          = "{Note on Hawking-Unruh effects in graphene}",
      journal        = "Mod. Phys. Lett.",
      volume         = "A27",
      year           = "2012",
      pages          = "1250218",
      doi            = "10.1142/S0217732312502185",
      eprint         = "1205.4039",
      archivePrefix  = "arXiv",
      primaryClass   = "gr-qc",
      SLACcitation   = "%%CITATION = ARXIV:1205.4039;%%"
}

@article{Das:2010mk,
      author         = "Das, Sumit R. and Ghosh, Archisman and Oh, Jae-Hyuk and Shapere, Alfred D.",
      title          = "{On Dumb Holes and their Gravity Duals}",
      journal        = "JHEP",
      volume         = "04",
      year           = "2011",
      pages          = "030",
      doi            = "10.1007/JHEP04(2011)030",
      eprint         = "1011.3822",
      archivePrefix  = "arXiv",
      primaryClass   = "hep-th",
      reportNumber   = "UK-10-11",
      SLACcitation   = "%%CITATION = ARXIV:1011.3822;%%"
}

@ARTICLE{deNova:2018rld,
       author = {{Ram{\'o}n Mu{\~n}oz de Nova}, Juan and {Golubkov}, Katrine and
        {Kolobov}, Victor I. and {Steinhauer}, Jeff},
        title = "{Observation of thermal Hawking radiation at the Hawking temperature in an analogue black hole}",
      journal = {arXiv e-prints},
         year = 2018,
        month = Sep,
          eid = {arXiv:1809.00913},
        pages = {arXiv:1809.00913},
archivePrefix = {arXiv},
       eprint = {1809.00913},
 primaryClass = {gr-qc},
       adsurl = {https://ui.adsabs.harvard.edu/\#abs/2018arXiv180900913R}
}

@article{Dey:2016khw,
      author         = "Dey, Ramit and Liberati, Stefano and Turcati, Rodrigo",
      title          = "{AdS and dS black hole solutions in analogue gravity: The relativistic and nonrelativistic cases}",
      journal        = "Phys. Rev.",
      volume         = "D94",
      year           = "2016",
      number         = "10",
      pages          = "104068",
      doi            = "10.1103/PhysRevD.94.104068",
      eprint         = "1609.00824",
      archivePrefix  = "arXiv",
      primaryClass   = "gr-qc",
      SLACcitation   = "%%CITATION = ARXIV:1609.00824;%%"
}

@article{Euve:2015vml,
      author         = "Euv{\'e}, L.-P. and Michel, F. and Parentani, R. and Philbin, T. G. and Rousseaux, G.",
      title          = "{Observation of noise correlated by the Hawking effect in a water tank}",
      journal        = "Phys. Rev. Lett.",
      volume         = "117",
      year           = "2016",
      number         = "12",
      pages          = "121301",
      doi            = "10.1103/PhysRevLett.117.121301",
      eprint         = "1511.08145",
      archivePrefix  = "arXiv",
      primaryClass   = "physics.flu-dyn",
      SLACcitation   = "%%CITATION = ARXIV:1511.08145;%%"
}

@article{Fedichev:2003id,
      author         = "Fedichev, Petr O. and Fischer, Uwe R.",
      title          = "{Gibbons-Hawking effect in the sonic de Sitter space-time of an expanding Bose-Einstein-condensed gas}",
      journal        = "Phys. Rev. Lett.",
      volume         = "91",
      year           = "2003",
      pages          = "240407",
      doi            = "10.1103/PhysRevLett.91.240407",
      eprint         = "cond-mat/0304342",
      archivePrefix  = "arXiv",
      primaryClass   = "cond-mat",
      SLACcitation   = "%%CITATION = COND-MAT/0304342;%%"
}

@article{Hartnoll:2009sz,
      author         = "Hartnoll, Sean A.",
      title          = "{Lectures on holographic methods for condensed matter physics}",
      booktitle      = "{Strings, Supergravity and Gauge Theories. Proceedings,
                        CERN Winter School, CERN, Geneva, Switzerland, February
                        9-13 2009}",
      journal        = "Class. Quant. Grav.",
      volume         = "26",
      year           = "2009",
      pages          = "224002",
      doi            = "10.1088/0264-9381/26/22/224002",
      eprint         = "0903.3246",
      archivePrefix  = "arXiv",
      primaryClass   = "hep-th",
      SLACcitation   = "%%CITATION = ARXIV:0903.3246;%%"
}

@article{Herzog:2011ec,
      author         = "Herzog, Christopher P. and Lisker, Nir and Surowka, Piotr and Yarom, Amos",
      title          = "{Transport in holographic superfluids}",
      journal        = "JHEP",
      volume         = "08",
      year           = "2011",
      pages          = "052",
      doi            = "10.1007/JHEP08(2011)052",
      eprint         = "1101.3330",
      archivePrefix  = "arXiv",
      primaryClass   = "hep-th",
      reportNumber   = "PUPT-2363",
      SLACcitation   = "%%CITATION = ARXIV:1101.3330;%%"
}

@article{Hossenfelder:2014gwa,
      author         = "Hossenfelder, Sabine",
      title          = "{Analog Systems for Gravity Duals}",
      journal        = "Phys. Rev.",
      volume         = "D91",
      year           = "2015",
      number         = "12",
      pages          = "124064",
      doi            = "10.1103/PhysRevD.91.124064",
      eprint         = "1412.4220",
      archivePrefix  = "arXiv",
      primaryClass   = "gr-qc",
      SLACcitation   = "%%CITATION = ARXIV:1412.4220;%%"
}

@article{Hossenfelder:2015pza,
      author         = "Hossenfelder, Sabine",
      title          = "{A relativistic acoustic metric for planar black holes}",
      journal        = "Phys. Lett.",
      volume         = "B752",
      year           = "2016",
      pages          = "13-17",
      doi            = "10.1016/j.physletb.2015.11.026",
      eprint         = "1508.00732",
      archivePrefix  = "arXiv",
      primaryClass   = "gr-qc",
      SLACcitation   = "%%CITATION = ARXIV:1508.00732;%%"
}

@article{Hossenfelder:2017iom,
      author         = "Hossenfelder, Sabine and Zingg, Tobias",
      title          = "{Analogue Gravity Models From Conformal Rescaling}",
      journal        = "Class. Quant. Grav.",
      volume         = "34",
      year           = "2017",
      number         = "16",
      pages          = "165004",
      doi            = "10.1088/1361-6382/aa7e12",
      eprint         = "1703.04462",
      archivePrefix  = "arXiv",
      primaryClass   = "gr-qc",
      SLACcitation   = "%%CITATION = ARXIV:1703.04462;%%"
}

@article{Hossenfelder:2018lhy,
    author = "Hossenfelder, Sabine and Zingg, Tobias",
    title = "{Analog Models for Holographic Transport}",
    eprint = "1810.05464",
    archivePrefix = "arXiv",
    primaryClass = "gr-qc",
    doi = "10.1103/PhysRevD.100.056015",
    journal = "Phys. Rev. D",
    volume = "100",
    number = "5",
    pages = "056015",
    year = "2019"
}

@article{Khveshchenko:2013foa,
      author         = "Khveshchenko, D. V.",
      title          = "{Simulating analogue holography in flexible Dirac metals}",
      journal        = "Europhys. Lett.",
      volume         = "104",
      year           = "2013",
      pages          = "47002",
      doi            = "10.1209/0295-5075/104/47002",
      eprint         = "1305.6651",
      archivePrefix  = "arXiv",
      primaryClass   = "cond-mat.str-el",
      SLACcitation   = "%%CITATION = ARXIV:1305.6651;%%"
}

@article{Peloquin:2015rnl,
      author         = "Peloquin, C{\'e}dric and Euv{\'e}, L{\'e}o-Paul and Philbin, Thomas and Rousseaux, Germain",
      title          = "{Analog wormholes and black hole laser effects in hydrodynamics}",
      journal        = "Phys. Rev.",
      volume         = "D93",
      year           = "2016",
      number         = "8",
      pages          = "084032",
      doi            = "10.1103/PhysRevD.93.084032",
      eprint         = "1512.03386",
      archivePrefix  = "arXiv",
      primaryClass   = "physics.flu-dyn",
      SLACcitation   = "%%CITATION = ARXIV:1512.03386;%%"
}

@article{Semenoff:2012xu,
      author         = "Semenoff, Gordon W.",
      title          = "{Engineering holographic graphene}",
      booktitle      = "{Proceedings, 6th International School on Field Theory
                        and Gravitation (ISFTG 2012): Petropolis, Rio de Janeiro,
                        Brazil, April 23-27, 2012}",
      journal        = "AIP Conf. Proc.",
      volume         = "1483",
      year           = "2012",
      pages          = "305-329",
      doi            = "10.1063/1.4756976",
      SLACcitation   = "%%CITATION = APCPC,1483,305;%%"
}

@article{Steinhauer:2014dra,
      author         = "Steinhauer, Jeff",
      title          = "{Observation of self-amplifying Hawking radiation in an analog black hole laser}",
      journal        = "Nature Phys.",
      volume         = "10",
      year           = "2014",
      pages          = "864",
      doi            = "10.1038/NPHYS3104",
      eprint         = "1409.6550",
      archivePrefix  = "arXiv",
      primaryClass   = "cond-mat.quant-gas",
      SLACcitation   = "%%CITATION = ARXIV:1409.6550;%%"
}

@article{Steinhauer:2015saa,
      author         = "Steinhauer, Jeff",
      title          = "{Observation of quantum Hawking radiation and its entanglement in an analogue black hole}",
      journal        = "Nature Phys.",
      volume         = "12",
      year           = "2016",
      pages          = "959",
      doi            = "10.1038/nphys3863",
      eprint         = "1510.00621",
      archivePrefix  = "arXiv",
      primaryClass   = "gr-qc",
      SLACcitation   = "%%CITATION = ARXIV:1510.00621;%%"
}

@article{Weinfurtner:2010nu,
      author         = "Weinfurtner, Silke and Tedford, Edmund W. and Penrice, Matthew C. J. and Unruh, William G. and Lawrence, Gregory A.",
      title          = "{Measurement of stimulated Hawking emission in an
                        analogue system}",
      journal        = "Phys. Rev. Lett.",
      volume         = "106",
      year           = "2011",
      pages          = "021302",
      doi            = "10.1103/PhysRevLett.106.021302",
      eprint         = "1008.1911",
      archivePrefix  = "arXiv",
      primaryClass   = "gr-qc",
      SLACcitation   = "%%CITATION = ARXIV:1008.1911;%%"
}

@article{Unruh:1980cg,
      author         = "Unruh, W. G.",
      title          = "{Experimental black hole evaporation}",
      journal        = "Phys. Rev. Lett.",
      volume         = "46",
      year           = "1981",
      pages          = "1351-1353",
      doi            = "10.1103/PhysRevLett.46.1351",
      SLACcitation   = "%%CITATION = PRLTA,46,1351;%%"
}

@article{Visser:1997ux,
      author         = "Visser, Matt",
      title          = "{Acoustic black holes: Horizons, ergospheres, and Hawking
                        radiation}",
      journal        = "Class. Quant. Grav.",
      volume         = "15",
      year           = "1998",
      pages          = "1767-1791",
      doi            = "10.1088/0264-9381/15/6/024",
      eprint         = "gr-qc/9712010",
      archivePrefix  = "arXiv",
      primaryClass   = "gr-qc",
      SLACcitation   = "%%CITATION = GR-QC/9712010;%%"
}

@article{Bilic:2022psx,
    author = "Bilic, Neven and Fabris, Julio C.",
    title = "{Thermodynamics of AdS planar black holes and holography}",
    eprint = "2208.00711",
    archivePrefix = "arXiv",
    primaryClass = "hep-th",
    reportNumber = "RBI-ThPhys-2022-39",
    doi = "10.1007/JHEP11(2022)013",
    journal = "JHEP",
    volume = "11",
    pages = "013",
    year = "2022"
}

@article{randall1,
    author = "Randall, Lisa and Sundrum, Raman",
    title = "{A Large mass hierarchy from a small extra dimension}",
    eprint = "hep-ph/9905221",
    archivePrefix = "arXiv",
    reportNumber = "MIT-CTP-2860, PUPT-1860, BUHEP-99-9",
    doi = "10.1103/PhysRevLett.83.3370",
    journal = "Phys. Rev. Lett.",
    volume = "83",
    pages = "3370--3373",
    year = "1999"
}

@article{randall2,
    author = "Randall, Lisa and Sundrum, Raman",
    title = "{An Alternative to compactification}",
    eprint = "hep-th/9906064",
    archivePrefix = "arXiv",
    reportNumber = "MIT-CTP-2874, PUPT-1867, BUHEP-99-13",
    doi = "10.1103/PhysRevLett.83.4690",
    journal = "Phys. Rev. Lett.",
    volume = "83",
    pages = "4690--4693",
    year = "1999"
}

@article{bombelli,
    author = "Bombelli, Luca and Koul, Rabinder K. and Lee, Joohan and Sorkin, Rafael D.",
    title = "{A Quantum Source of Entropy for Black Holes}",
    reportNumber = "PRINT-86-0371 (SYRACUSE)",
    doi = "10.1103/PhysRevD.34.373",
    journal = "Phys. Rev. D",
    volume = "34",
    pages = "373--383",
    year = "1986"
}

@article{srednicki,
    author = "Srednicki, Mark",
    title = "{Entropy and area}",
    eprint = "hep-th/9303048",
    archivePrefix = "arXiv",
    reportNumber = "LBL-33754, CFPA-93-02",
    doi = "10.1103/PhysRevLett.71.666",
    journal = "Phys. Rev. Lett.",
    volume = "71",
    pages = "666--669",
    year = "1993"
}

@article{uhlmann,
    author = "Uhlmann, Michael and Xu, Yan and Schutzhold, Ralf",
    title = "{Aspects of cosmic inflation in expanding Bose-Einstein condensates}",
    eprint = "quant-ph/0509063",
    archivePrefix = "arXiv",
    doi = "10.1088/1367-2630/7/1/248",
    journal = "New J. Phys.",
    volume = "7",
    pages = "248",
    year = "2005"
}

@article{girelli,
    author = "Girelli, Florian and Liberati, Stefano and Sindoni, Lorenzo",
    title = "{Gravitational dynamics in Bose Einstein condensates}",
    eprint = "0807.4910",
    archivePrefix = "arXiv",
    primaryClass = "gr-qc",
    doi = "10.1103/PhysRevD.78.084013",
    journal = "Phys. Rev. D",
    volume = "78",
    pages = "084013",
    year = "2008"
}

@article{fleurov,
  title={Regularization of fluctuations near the sonic horizon due to the quantum potential and its influence on Hawking radiation},
  author={Fleurov, V and Schilling, R},
  journal={Physical Review A—Atomic, Molecular, and Optical Physics},
  volume={85},
  number={4},
  pages={045602},
  year={2012},
  publisher={APS}
}

@article{rinaldi,
    author = "Rinaldi, Massimiliano",
    title = "{The entropy of an acoustic black hole in Bose-Einstein condensates}",
    eprint = "1106.4764",
    archivePrefix = "arXiv",
    primaryClass = "gr-qc",
    doi = "10.1103/PhysRevD.84.124009",
    journal = "Phys. Rev. D",
    volume = "84",
    pages = "124009",
    year = "2011"
}

@article{anderson,
    author = "Anderson, Paul R. and Balbinot, Roberto and Fabbri, Alessandro and Parentani, Renaud",
    title = "{Hawking radiation correlations in Bose Einstein condensates using quantum field theory in curved space}",
    eprint = "1301.2081",
    archivePrefix = "arXiv",
    primaryClass = "gr-qc",
    doi = "10.1103/PhysRevD.87.124018",
    journal = "Phys. Rev. D",
    volume = "87",
    number = "12",
    pages = "124018",
    year = "2013"
}

@article{ryu,
    author = "Ryu, Shinsei and Takayanagi, Tadashi",
    title = "{Holographic derivation of entanglement entropy from AdS/CFT}",
    eprint = "hep-th/0603001",
    archivePrefix = "arXiv",
    reportNumber = "NSF-KITP-06-11, NSF-KITP-06-11",
    doi = "10.1103/PhysRevLett.96.181602",
    journal = "Phys. Rev. Lett.",
    volume = "96",
    pages = "181602",
    year = "2006"
}

@article{ryu2,
    author = "Ryu, Shinsei and Takayanagi, Tadashi",
    title = "{Aspects of Holographic Entanglement Entropy}",
    eprint = "hep-th/0605073",
    archivePrefix = "arXiv",
    reportNumber = "NSF-KITP-06-31, KUNS-2021, NSF-KITP-06-31, KUNS-2021",
    doi = "10.1088/1126-6708/2006/08/045",
    journal = "JHEP",
    volume = "08",
    pages = "045",
    year = "2006"
}

@article{bobev,
    author = "Bobev, Nikolay and Kundu, Arnab and Pilch, Krzysztof and Warner, Nicholas P.",
    title = "{Minimal Holographic Superconductors from Maximal Supergravity}",
    eprint = "1110.3454",
    archivePrefix = "arXiv",
    primaryClass = "hep-th",
    reportNumber = "UTTG-10-11",
    doi = "10.1007/JHEP03(2012)064",
    journal = "JHEP",
    volume = "03",
    pages = "064",
    year = "2012"
}

@article{albash,
    author = "Albash, Tameem and Johnson, Clifford V.",
    title = "{Holographic Studies of Entanglement Entropy in Superconductors}",
    eprint = "1202.2605",
    archivePrefix = "arXiv",
    primaryClass = "hep-th",
    doi = "10.1007/JHEP05(2012)079",
    journal = "JHEP",
    volume = "05",
    pages = "079",
    year = "2012"
}

@article{bilic4,
    author = "Bili{\'c}, Neven and Fabris, Julio C.",
    title = "{Analog dual to a 2 + 1-dimensional holographic superconductor}",
    eprint = "2101.12494",
    archivePrefix = "arXiv",
    primaryClass = "hep-th",
    reportNumber = "RBI-ThPhys-2021-7",
    doi = "10.1088/1361-6382/ac1207",
    journal = "Class. Quant. Grav.",
    volume = "38",
    number = "16",
    pages = "165007",
    year = "2021"
}

@article{lemos1,
    author = "Lemos, Jose P. S.",
    title = "{Two-dimensional black holes and planar general relativity}",
    eprint = "gr-qc/9407024",
    archivePrefix = "arXiv",
    doi = "10.1088/0264-9381/12/4/014",
    journal = "Class. Quant. Grav.",
    volume = "12",
    pages = "1081--1086",
    year = "1995"
}

@article{witten2,
    author = "Witten, Edward",
    editor = "Bergstrom, L. and Lindstrom, U.",
    title = "{Anti-de Sitter space, thermal phase transition, and confinement in gauge theories}",
    eprint = "hep-th/9803131",
    archivePrefix = "arXiv",
    reportNumber = "IASSNS-HEP-98-21",
    doi = "10.4310/ATMP.1998.v2.n3.a3",
    journal = "Adv. Theor. Math. Phys.",
    volume = "2",
    pages = "505--532",
    year = "1998"
}

@article{bilic2,
    author = "Bilic, Neven",
    editor = "Buric, Maja and Djordjevic, Goran and Haack, Michael and Lust, Dieter and Senjanovic, Goran",
    title = "{Thermodynamics of dark energy}",
    eprint = "0812.5050",
    archivePrefix = "arXiv",
    primaryClass = "gr-qc",
    doi = "10.1002/prop.200710507",
    journal = "Fortsch. Phys.",
    volume = "56",
    pages = "363--372",
    year = "2008"
}

@article{bilic3,
    author = "Bilic, Neven and Tolic, Dijana",
    title = "{FRW universe in the laboratory}",
    eprint = "1309.2833",
    archivePrefix = "arXiv",
    primaryClass = "gr-qc",
    doi = "10.1103/PhysRevD.88.105002",
    journal = "Phys. Rev. D",
    volume = "88",
    pages = "105002",
    year = "2013"
}
\bibliographystyle{abbrv}

\end{document}